\documentclass[preprint,prX,showpacs]{revtex4} %PRL format
\usepackage{epsfig}
\usepackage{amssymb,amsmath,amsbsy,wasysym}
\usepackage[normalem]{ulem} % for strike out emphasis

\newcommand{\beq}{\begin{equation}}
\newcommand{\eeq}{\end{equation}}
\newcommand{\beqarr}{\begin{eqnarray}}
\newcommand{\eeqarr}{\end{eqnarray}}

\begin{document}
%opening
\title{Perturbation Analysis of Complete Synchronization in Networks of Phase Oscillators}
\author{Ralf T{\"o}njes$^{1,2}$}
\author{Bernd Blasius$^{3}$}
\affiliation{${}^{1}$ Institut f\"ur Physik, Universit\"at Potsdam, 14415 Potsdam, Germany}
\affiliation{${}^{2}$ Ochadai Academic Production, Ochanomizu University, Tokyo 112-8610, Japan}
\affiliation{${}^{3}$ ICBM, University Oldenburg, 26111 Oldenburg, Germany}

\begin{abstract}
\noindent
The behavior of weakly coupled self-sustained oscillators can often be well described by phase equations. Here we use the paradigm of Kuramoto phase oscillators which are coupled in a network to calculate first and second order corrections to the frequency of the fully synchronized state for nonidentical oscillators. The topology of the underlying coupling network is reflected in the eigenvalues and eigenvectors of the network Laplacian which influence the synchronization frequency in a particular way. They characterize the importance of nodes in a network and the relations between them. Expected values for the synchronization frequency are obtained for oscillators with quenched random frequencies on a class of scale-free random networks and for a Erd\H{o}s-R\'enyi random network. We briefly discuss an application of the perturbation theory in the second order to network structural analysis.
\end{abstract}
\pacs{05.45.Xt, 64.60.aq}

\maketitle

\section*{I. Introduction}
\noindent
The collective behavior of ensembles of interacting units is one of the main topics in complex system theory. Different parts of a complex system can be identified as subsystems and studied individually, while the interaction between these can lead to emergent properties of the whole system. In particular synchronization, the adjustment of internal timescales in oscillatory systems which interact locally or through a complex network \cite{PiRoKurths03,OsChanKu07} , is ubiquitous in biological \cite{Winfree67,Tass99,ErmKop84,BlasiusStone99,KoriMikh04,KoriMikh06} and technical applications \cite{WiesColStro96,SiFaWie93,LaKoPeHe06,DiAre08}. Recently also chemical reactions with feedback control have been proposed to realize specific interaction topologies \cite{BlasiusShow07,Show05}. Synchronization can orchestrate macroscopic spatio-temporal periodicity even if the individual units are very different from each other and a simple linear superposition of their output would be incoherent. While this is a desirable effect in many applications, such as coupled Josephson junctions or laser arrays \cite{WiesColStro96,SiFaWie93} it can also lead to pathological states like epilepsy or Parkinson's disease \cite{Tass99} or it can be disastrous when it occurs in constructions \cite{OttStrogatz05}. 
\\ \\
The onset of synchronization for very heterogeneous systems has been described as a second order phase transition in the limit of large system sizes \cite{Kuramoto75,Kuramoto84}. Above a critical coupling strength or below a critical heterogeneity the incoherent state becomes unstable and global collective behavior can be observed \cite{Kuramoto84,RespOtt05,Rosenblum07}. For identical, possibly chaotic, subsystems complete synchronization can be possible \cite{FujiYa83,PecoCar98}. It is known that the spectral properties of the coupling network play an important role in the transition to synchronization \cite{OsChanKu07,RespOtt05} and the stability of complete synchronization \cite{FujiYa83,PecoCar98}. But many studies on synchronization in networks have mainly been concerned with the estimation of a few important eigenvalues of the network Laplacian \cite{ChanMottKu05}.
\\ \\
In this paper we study the synchronization frequency in networks of weakly nonidentical, autonomous oscillators with attractive coupling. Under these conditions the Kuramoto phase equations (KPE) \cite{Kuramoto84} can be used to describe the system qualitatively and quantitatively. The KPE show a rich collective behavior with transitions from complete desynchronization, where the phases are uniformly distributed, to partial synchronization with a unimodal distribution of phases or even clustering \cite{Kuramoto84,RespOtt05,Rosenblum07,ErmChimera08,KoriKiss07,KoriKiss08} and finally 
%
%phase synchronization, also known as 
frequency synchronization or phase locking, where the phase difference between any two oscillators is constant. 
\\ \\
%
%Complete synchronization is a stable solution of the KPE in systems of identical phase oscillators with attractive coupling.
In systems of identical phase oscillators with attractive coupling complete synchronization is a stable solution of the KPE.
We will quantify the frequency heterogeneity of the oscillators and derive a perturbation expansion around the well known synchronization manifold for identical oscillators in powers of the heterogeneity. In analogy to perturbation theory for the continuous, nonlinear Kuramoto Phase Diffusion Equation \cite{BlaToe08},  in Sections II and III we will show two approaches which lead to the same first and second order perturbation terms. In random networks, the expected second order perturbation term of the synchronization frequency can be interpreted as a mean value with respect to the spectral density of the network Laplacian. Using a random network model for which the spectral density of the Laplacian is known, we explicitly calculate the expected synchronization frequency in Section IV. We verify out theory by numerical simulations.
\subsection*{The Kuramoto model}
\noindent
Let us briefly review the Kuramoto Phase Equations (KPE) for discretely coupled oscillators \cite{Kuramoto75,Kuramoto84}. The dynamics of an ensemble of $N$ autonomous oscillators may be given as
\beq
	\dot{\mathbf{X}}_n = \mathbf{F}_n(\mathbf{X}_n) + \sum_{m=1}^N \mathbf{V}_{nm}(\mathbf{X}_m,\mathbf{X}_n) ~,
\eeq
where $\mathbf{X}_n$ defines the state of the oscillator labeled with $n$, the velocity field $\mathbf{F}_n$ allows for stable limit cycle oscillations and $\mathbf{V}_{nm}$ describes the coupling between two oscillators depending on their state. 
In his monograph \cite{Kuramoto84} Kuramoto considered the heterogeneity in the oscillators as well as the coupling as a perturbation of a common oscillator dynamics $\mathbf{F}_n=\mathbf{F}+\delta\mathbf{F}_n$. For this common dynamics one can define a uniformly evolving phase variable $\phi$ in a neighborhood of the limit cycle. The dynamics of the phases $\phi_n$ in linear response to the perturbation corresponds to the phase model introduced by Winfree \cite{Winfree67}
\beq
%	{\dot\phi}_n = \omega_n + \sum_{m=1}^N \mathbf{Z}_n(\phi_n){~}\mathbf{V}_{nm}(\phi_m,\phi_n) ~.
	{\dot\phi}_n = \omega + \delta\omega_n(\phi_n) + \sum_{m=1}^N \mathbf{Z}^\dagger(\phi_n){~}\mathbf{V}_{nm}(\phi_m,\phi_n) ~.
\eeq
Here $\omega$ is the natural frequency and $\mathbf{Z}$ is called the phase response function of 
%the oscillator $n$. 
the common oscillator dynamics.
If the phase differences change slowly over the time of one oscillation then one can use phase averaging techniques \cite{Kuramoto84,PiRoKurths03} to obtain effective phases $\vartheta_n$ and phase equations which only depend on the phase differences. If we finally assume that the functional form of the coupling between any two oscillators $n$ and $m$ only differs in a coupling constant $A_{nm}$ we obtain the KPEs
\beq	\label{Eq:KPE01}
	{\dot\vartheta}_n = \omega_n + \sum_{m=1}^N A_{nm}~g(\vartheta_m-\vartheta_n)		~.
\eeq
The phase coupling function $g(\Delta\vartheta)$ is periodic. 
%For identical oscillators and diffusive coupling it vanishes at zero phase differences and has a positive derivative at zero. It can then be approximated by its lowest Fourier modes as
For diffusive coupling it vanishes at zero. We assume a positive derivative at zero and approximate the coupling function by its lowest Fourier modes as
\beq	\label{Eq:CoupFunc01}
	g(\Delta\vartheta) = \sin\Delta\vartheta + \gamma \left( 1 - \cos\Delta\vartheta \right)	~.
\eeq
The parameter $\gamma$ breaks the symmetry of the phase coupling function and can directly be associated with the amplitude dependence of the phase velocity in complex Stuart-Landau oscillators \cite{Kuramoto84}, i.e. a third-order nonlinear effect in the normal form of a supercritical Hopf bifurcation also known as nonisochronicity. The effect of nonisochronicity on the ability of a system to synchronize and on the formation of spatio-temporal patterns has been noted early on \cite{Sakaguchi88} and again stressed recently \cite{ErmChimera08,BlaToe05,Blasius05} whereas it is often disregarded in favor of analytic simplicity \cite{Kuramoto84,KoriMikh04,RespOtt05}.
\\ \\
%Phase synchronization 
A fully phase locked state
is reached when the oscillators can arrange their phases in a way that due to an exact balance of nonidentical natural frequencies and coupling forces all oscillators have the same synchronization frequency
\beq	\label{Eq:KPESynch01}
	\Omega = \sigma\eta_n + \sum_{m=1}^N A_{nm}~g(\vartheta_m-\vartheta_n)	~.
\eeq
The frequencies $\eta_n$ in this equation are normalized to have unit variance.
Then the heterogeneity of the oscillators is quantified by the variance $\textnormal{var}(\omega)=\sigma^2$ of the natural frequencies in the system. The mean frequency $\bar\omega$ does not necessarily depend on the heterogeneity $\sigma$ but here we choose a co-rotating frame of reference where $\bar\omega=\sigma\bar\eta$. For identical oscillators ($\sigma=0$) complete synchronization with identical phases $\vartheta^{(0)}_n=\vartheta^{(0)}_m$ for all $n$ and $m$, and synchronization frequency $\Omega^{(0)}=0$ is a solution of Eq.(\ref{Eq:KPESynch01}) with
\beq
	\Omega^{(0)} ~=~ 0 ~=~ \sum_{m=1}^N A_{nm}~g(\vartheta^{(0)}_m - \vartheta^{(0)}_n)	~.
\eeq
\subsection*{Stability of the synchronized state}
\noindent
Under some weak conditions on the coupling topology one can show that the state of complete synchronization is stable. But it has been shown recently, that in a heterogeneous coupling network and for large nonisochronicity $\gamma$ the stable state of complete synchronization can co-exist with a dynamical equilibrium of complete desynchronization or partial synchronization 
\cite{ErmChimera08}.  Conversely, if the nonisochronicity is not too high and the network is well connected, complete synchronization is the typical result from random initial conditions.
\\ \\
A sufficient condition for the stability of complete synchronization of identical oscillators is that all values $A_{nm}$ are non-negative and the corresponding weighted network is strongly connected, i.e. there exists a path %with positive link  weights 
between any two nodes. To see this, one can consider small deviations $\varphi_n$ from the synchronized solution. Linearizing Eq. (\ref{Eq:KPE01}) for $\sigma=0$ and small deviations around $\boldsymbol{\vartheta}^{(0)}$ one obtains
\beq
	\dot{\varphi}_n = \sum_{m=1}^N A_{nm}~\left(\varphi_m-\varphi_n\right) = \sum_{m=1}^N L_{nm}~\varphi_m	~,
\eeq
with the network Laplacian $\mathbf{L}$ defined as
\beq	\label{Eq:LapDef}
	L_{nm} = A_{nm} - \delta_{nm} \sum_{l=1}^N A_{nl}	~.
\eeq
Since all row sums $\sum_m L_{nm}$ are zero at least one eigenvalue $\lambda_0$ of the network Laplacian is also zero, corresponding to a constant shift of all phases along the synchronization manifold.
If all values $A_{nm}$ are non-negative then it follows from the Gershgorin circle theorem that the network Laplacian has only non-positive eigenvalue real parts $0 = \lambda_0 \ge \textnormal{Re}\lambda_1 \ge \dots \ge \textnormal{Re}\lambda_{N-1}$, where $N$ is the number of oscillators. Complete synchronization is unique up to a global phase shift, only if the second largest eigenvalue real part  $\textnormal{Re}\lambda_1$ is strictly smaller than zero. 
\\ \\
Associated with the relaxation to synchronization is a diffusion process in the opposite direction of the coupling. If all off-diagonal elements are non-negative, the transposed Laplacian $\mathbf{L}^\dagger$ can be viewed as a matrix of transition rates for a master equation $\dot{\mathbf P}=\mathbf{L}^\dagger\mathbf{P}$ with a probability vector $\mathbf{P}$. The eigenvalue $\lambda_0$ is non-degenerate if and only if the stationary probability distribution $\mathbf{P}_0$ is unique, i.e. independent of the initial condition. Note, that a strongly connected network of transition rates is sufficient but not necessary for that \cite{VanKampen81}. In the following we will assume that $\textnormal{Re}\lambda_k<0$ for all $k>0$.
\section*{II. Perturbation Approach 1}
\noindent
The algebraic equation Eq.(\ref{Eq:KPESynch01}) implicitly defines the synchronization frequency and the phases in synchronization (up to global phase shift), even for non-zero heterogeneity. However, a stable phase locked solution 
of Eq.(\ref{Eq:KPESynch01})
or any solution at all may not exist. Only for small heterogeneity we can expect that a stable solution exists, that it is close to the solution for identical oscillators ($\sigma=0$) and that it can be expanded in powers of $\sigma$ as
\beqarr	\label{Eq:PertAnsatz01}
	\boldsymbol{\vartheta} &=& \boldsymbol{\vartheta}^{(0)} + \sigma\boldsymbol{\vartheta}^{(1)} + \sigma^2\boldsymbol{\vartheta}^{(2)} + ~O(\sigma^3)	~, \nonumber \\ \\
	\Omega &=& \sigma\Omega^{(1)} + \sigma^2\Omega^{(2)} + ~O(\sigma^3) 	~.\nonumber
\eeqarr
In this section we follow closely the procedure outlined in \cite{BlaToe08} to derive the perturbation expansion of the Kuramoto Phase Equations in synchronization Eq.(\ref{Eq:KPESynch01}). We directly insert the ansatz Eq.(\ref{Eq:PertAnsatz01}) into Eq.(\ref{Eq:KPESynch01}), use a Taylor expansion of the coupling function around zero and regroup the terms according to powers of $\sigma$. This procedure requires sorting of infinite summations and some careful consideration of the index limits. It is shown in detail in the Appendix A. However, the result takes a simple form in vector notation
\beq	\label{Eq:OhmExpansionSolved}
	\Omega^{(l)}\mathbf{1} = \left(\mathbf{L}\boldsymbol{\vartheta}^{(l)} + \mathbf{b}^{(l)}\right)	~.
\eeq
Here $\Omega^{(l)}$ is the $l^\textnormal{th}$ order perturbation term of $\Omega$ in Eq.(\ref{Eq:PertAnsatz01}), the vector $\boldsymbol{\vartheta}^{(l)}$ is the corresponding perturbation term for the phases in synchronization, $\mathbf{1}$ is a constant vector with unity entries, the matrix $\mathbf{L}$ is the Laplacian of the network, as defined in Equation Eq.(\ref{Eq:LapDef}) and $\mathbf{b}^{(l)}$ is a vector which depends nonlinearly on all perturbation terms of order lower than $l$ (see Eqs.~(\ref{Eq:bI})-(\ref{Eq:bIII})). Equation (\ref{Eq:OhmExpansionSolved}) can thus be solved iteratively for each order of perturbation. In practice, while the amplitude of the terms $\mathbf{b}^{(l)}$ is as small as $O(\sigma^l)$, the analytic expression and the expense for its calculation blows up quickly.
\\ \\
Let us consider a complete, orthonormal set of left and right eigenvectors $\mathbf{P}_k$ and $\mathbf{p}_k$ of the network Laplacian with
\beqarr	\label{Eq:EigenSystem}
	\mathbf{L}\mathbf{p}_k = \lambda_k \mathbf{p}_k ~,&&\quad \mathbf{L}^\dagger\mathbf{P}_k = \lambda_k^*\mathbf{P}_k ~,\nonumber \\ \\
	\mathbf{P}_k^\dagger{~}\mathbf{p}_{k'} = \delta_{kk'} ~,&&\quad \sum_{k=0}^{N-1} \mathbf{p}_k\mathbf{P}_k^\dagger = \mathbb{I}	~,\nonumber
\eeqarr
and in particular
\beq
	\mathbf{p}_0 = \mathbf{1}	\quad\textnormal{and}\quad		\mathbf{1}^\dagger{~}\mathbf{P}_0 = 1		~.
\eeq
The left eigenvector $\mathbf{P}_0$, which is the stationary solution of the master equation $\dot{\mathbf{P}}=\mathbf{L}^\dagger\mathbf{P}$, assigns a weight to each node of the network \cite{KoriAraiKura09}. The solution of Equation (\ref{Eq:OhmExpansionSolved}) is
\beqarr \label{Eq:PertSol01}
	\Omega^{(l)}  &=& \mathbf{P}_0^\dagger{~}\mathbf{b}^{(l)} ~,%~=~ \left\langle b^{(l)}\right\rangle_{\mathbf{P}_0}	
\nonumber \\ \\
	\boldsymbol{\vartheta}^{(l)} &=& -\sum_{k\ne 0} \frac{ \left(\mathbf{P}_k^\dagger{~}\mathbf{b}^{(l)}\right) }{\lambda_k} \mathbf{p}_k	\nonumber ~.
\eeqarr
Using the short notations $g''_0=g''(0)$, $g'''_0=g'''(0)$ and $\vartheta_{mn}^{(l)} = \vartheta_m^{(l)} - \vartheta_n^{(l)}$
the first three vectors $\mathbf{b}^{(1)}$, $\mathbf{b}^{(2)}$ and $\mathbf{b}^{(3)}$ are
\beqarr
	b_n^{(1)} &=& \eta_n	~,\label{Eq:bI}	\\ \nonumber \\
	b_n^{(2)} &=& \sum_{m=1}^N A_{nm} \frac{1}{2}g''_0{\vartheta_{mn}^{(1)}}^2	~,	\label{Eq:bII}	\\ \nonumber \\
	b_n^{(3)} &=& \sum_{m=1}^N A_{nm} \left( g''_0\vartheta_{mn}^{(1)}\vartheta_{mn}^{(2)} + \frac{1}{6}g'''_0{\vartheta_{mn}^{(1)}}^3 \right)	\label{Eq:bIII} ~.
\eeqarr
Equations (\ref{Eq:PertSol01})-(\ref{Eq:bIII}) give the first three perturbation terms of the synchronization frequency and the relative phases in synchronization. In the next section, we will derive the first and the second order terms again, but in a slightly different form which allows for a better analysis.
\section*{III. Perturbation Approach 2}
\noindent
The second order correction Eqs.~(\ref{Eq:PertSol01}) and (\ref{Eq:bII}) of the synchronization frequency depends on the second derivative $g''_0=g''(0)$ of the phase coupling function at zero. If we are only interested in the first and second order perturbation terms we have the freedom to choose a different coupling function $\tilde{g}(\varphi)$ in the equation Eq.(\ref{Eq:KPESynch01}) with the same first and second derivative at zero as $g(\varphi)$ which may be more suitable for an analysis. For the continuous Kuramoto Phase Diffusion equations it is known that a non-linear Cole-Hopf transformation $\vartheta=\gamma^{-1}\log p$ changes the equations in synchronization into an eigenvalue problem of a stationary, linear Schr\"odinger equation \cite{Kuramoto84,Sakaguchi88,BlaToe05,BlaToe08}. With the same procedure in mind we will define an auxiliary coupling function $\tilde{g}(\varphi)$ as
\beq	\label{Eq:CoupFunc02}
	\tilde{g}(\varphi) ~=~ \frac{1}{\gamma}\left(e^{\gamma\varphi}-1\right) ~=~ g(\varphi) ~+~ O(\varphi^3)	~.
\eeq
After the transformation
\beq
	\vartheta_n = \frac{1}{\gamma}\log p_n	~,
\eeq
we can bring equation Eq.(\ref{Eq:KPESynch01}) with $\tilde{g}(\varphi)$ as coupling function into the form of an eigenvalue problem
\beq
	-E_0 \mathbf{p} ~=~ \gamma\Omega \mathbf{p} ~=~ \bigl[\gamma\sigma \mathbf{V}_{\eta} + \mathbf{L}\bigr]~ \mathbf{p} ~=~ -\mathbf{H}\mathbf{p}	~,
\eeq
where the vector $\mathbf{p}$ has the entries $p_n$, $\mathbf{V}_{\eta}=\textnormal{diag}(\boldsymbol{\eta})$ is the diagonal matrix of frequencies and $\mathbf{L}$ is the network Laplacian (Eq.(\ref{Eq:LapDef})). This equation has the form of a stationary discrete Schr\"odinger equation for the ground state of a particle hopping between the vertices of the coupling graph with the on-site potentials $-\eta_n$ and ground state energy $E_0=-\gamma\Omega$. If the coupling network is symmetric the Hamiltonian $\mathbf{H}$ is symmetric, as well, the left and right eigenvectors are identical and all eigenvalues are real. In this section we will not yet make this simplifying assumption.
\\ \\
The potential $\mathbf{V}_{\eta}$ of random frequencies can be treated as a perturbation of strength $\gamma\sigma$ of the eigenvalue problem for the network Laplacian. Given the eigenvalues and orthonormal left and right eigenfunctions of $\mathbf{L}$ Eq.(\ref{Eq:EigenSystem}) we are looking for the coefficients $E_0^{(l)}$ of the expansion
\beq
	-E_0 ~=~ \lambda_0 ~-~ \gamma\sigma E_0^{(1)} ~-~ \gamma^2\sigma^2 E_0^{(2)} ~-~ O(\gamma^3\sigma^3)		~.
\eeq
Again, we assume that the eigenvalue $\lambda_0=0$ of the Laplacian is non-degenerate, so that the ground state is unique up to normalization. Accordingly, the synchronized state is unique up to a constant phase shift. Schr\"odinger perturbation theory, modified to allow for asymmetric operators gives the expressions
\beqarr
	-E_0^{(1)} &=& \left(\mathbf{P_0}^\dagger {~} \mathbf{V}_{\eta}\mathbf{p}_0\right) 	~,\nonumber \\ \\
	-E_0^{(2)} &=& \sum_{k\ne 0} \frac{	\left(\mathbf{P}_0^\dagger{~}\mathbf{V}_{\eta}\mathbf{p}_k\right)
						\left(\mathbf{P}_k^\dagger{~}\mathbf{V}_{\eta}\mathbf{p}_0\right)}{\lambda_0-\lambda_k}	~.	\nonumber
\eeqarr
For the first two coefficients in the perturbation expansion Eq.(\ref{Eq:PertAnsatz01}) of the synchronization frequency $\Omega$ we find
\beqarr	\label{Eq:PertSol02}
	\Omega^{(1)} 	&=& \mathbf{P}_0^\dagger{~}\boldsymbol{\eta} 	~,%\left\langle\eta\right\rangle_{\mathbf{P}_0}		
\nonumber \\ \\
	\Omega^{(2)}	&=& -\gamma\sum_{k\ne 0} \frac{	\left(\mathbf{P}_0^\dagger{~}\mathbf{V}_{\eta}\mathbf{p}_k\right)
						\left(\mathbf{P}_k^\dagger{~}\mathbf{V}_{\eta}\mathbf{p}_0\right)}{\lambda_k}	~.	\nonumber
\eeqarr
The first term is the weighted average of the frequencies with respect to the stationary probability distribution of the master equation $\dot{\mathbf{P}}=\mathbf{L}^\dagger\mathbf{P}$ with the transposed network Laplacian as matrix of transition rates. In the second expression we have used $\Omega^{(2)}=-\gamma E_0^{(2)}$ and $\lambda_0=0$. Equation Eq.(\ref{Eq:PertSol02}) is a more compact form  of equations Eqs.~(\ref{Eq:PertSol01})-(\ref{Eq:bII}) combined.
\section*{IV. Examples}
\noindent
We can now study the change of the synchronization frequency with respect to oscillator heterogeneity and to the architecture of the coupling network. To find expressions for the expected first and second order response we will consider the ensemble of different realizations of random frequencies and an ensemble of random networks. Let us assume independent, identically distributed random frequencies with $\mathbb{E}\left[\eta_n\eta_m\right] - \mathbb{E}\left[\eta\right]^2 = \delta_{nm}$ 
Then from Eq.(\ref{Eq:PertSol02}) follows
\beqarr
	\mathbb{E}\left[\Omega^{(1)}\right] 	&=& \mathbb{E}_\eta\left[\eta\right]	~,	\nonumber \\ \\
	\mathbb{E}\left[\Omega^{(2)}\right]	&=& -\gamma~ \mathbb{E}_{NW} \left[ \sum_{k\ne 0}
							\frac{\left(\mathbf{P}_k^\dagger{~}\mathbf{V}_{\mathbf{P}_0}~\mathbf{p}_k\right)}{\lambda_k}	\right]	~. \nonumber \label{Eq:SpecW2}
\eeqarr
Here $\mathbb{E}_\eta$ is the expected value with respect to the frequency distribution and $\mathbb{E}_{NW}$ over the network ensemble. The vectors $\mathbf{P}_k$ and $\mathbf{p}_k$ in the second equation are left and right eigenvectors of the network Laplacian and $\lambda_k\ne 0$ the corresponding eigenvalues. The operator $\mathbf{V}_{\mathbf{P}_0}$ is diagonal with the components of $\mathbf{P}_0$ on the diagonal. 
The first order frequency correction is independent of the network architecture while for uncorrelated frequencies the second order perturbation term is 
determined by the topology of the coupling network 
and the frequency heterogeneity $\sigma^2$. 
For symmetric coupling $A_{nm}=A_{mn}$ the left eigenvector $\mathbf{P}_0$ is given by $N^{-1}\mathbf{1}$ and in the limit $N\to\infty$ the expected value of $\Omega^{(2)}$ in Eq. (\ref{Eq:SpecW2}) can be written as
\beq	\label{Eq:SpecAv}
	\mathbb{E}\left[\Omega^{(2)}\right] ~=~ \gamma\int ~ \rho(\lambda) ~\frac{1}{\lambda}~d\lambda	~,
\eeq
where $\rho(\lambda)$ is the Laplacian spectral density of the random network ensemble.
\\ \\
The integral Eq.~(\ref{Eq:SpecAv}) has been studied in the different context of vibrational thermodynamic stability for networks of linear springs, modelling complex molecules \cite{BuCaFoVu02}. Whether the integral is finite depends on the spectral dimension $d$ of the network, defined by the limit behavior of $\rho(\lambda)\sim\lambda^{\frac{d}{2}-1}$ for $\lambda\to 0$, a suitable generalization of the Euclidean dimension for regular lattices to geometrically disordered structures \cite{OrAl82}. For networks with spectral dimension $d>2$ larger than two, the integral in Eq.~(\ref{Eq:SpecAv}) is finite. In \cite{BuCaFoVu02} the authors study the case of a Sierpinski gasket which is a fractal graph for which the Laplacian spectrum can be calculated analytically and the spectral dimension is lower than two. In \cite{BlaToe08} we study regular topologies for which the Fourier spectral decomposition is known and we also find the lower critical dimension $d=2$. For ensembles of random graphs in general, it is a complicated task to find analytic expressions for the spectral density. Approximations of the spectral density of matrices associated with complex random networks, such as the Wigner semicircle law, are usually only available in the limit of dense networks, where the mean degree is much larger than one. 
\begin{figure}[thb] 
\center
\includegraphics[width=12.5cm]{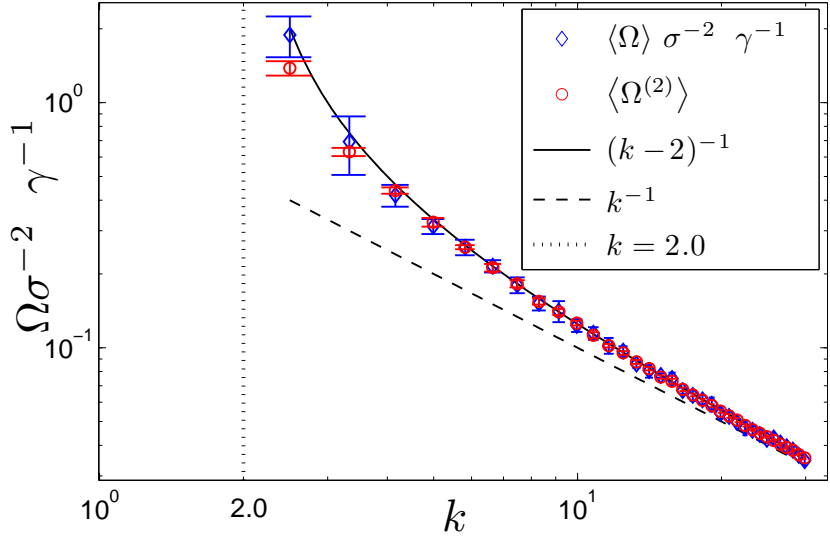}
\caption{\small (color online)
Frequency shift in undirected, random networks of size $N=400$ with Poissonian degree distribution and given mean degree $k$. The mean degree must be larger than $k_{cr}=2$ for an infinite random tree network. The frequency shift could be measured precisely by enforcing a zero mean frequency $\bar\omega=0$. Random frequencies were drawn from a uniform distribution of standard deviation $\sigma=10^{-1}$. From an ensemble of 10 realizations we show the mean network synchronization frequency divided by variance $\sigma^2$ and nonisochronicity $\gamma$ (blue diamonds, Newton Method, $\gamma=1.0$), the predicted second order term $\langle\Omega^{(2)}\rangle$ from equation Eq.(\ref{Eq:SpecW2}) and the spectrum of the network Laplacian (red circles), the asymptotic line $k^{-1}$ (dashed line) and the line $(k-2)^{-1}$ (solid line), which describes the actual behavior of the frequency shift even for small mean degrees close to $k_{cr}=2$.
}
\label{Fig:fig01}
\end{figure}
\subsection*{A static scale-free random network model}
\noindent
As an example we will use a recent result for the Laplacian of a static scale-free random network model \cite{Kahng01,Kahng07}. For this model the coupling strength $A_{nm}=A_{mn}$ between two oscillators is either zero or it is one with the probability $kNw_nw_m$, where the $w_n\sim n^{-1/(\alpha-2)}$ are normalized weights for the nodes $n=1\dots N$, and $k$ is the mean degree of the network. The degree distribution follows a power law with exponent $-\alpha$. In the thermodynamic limit $N\to\infty$ and large $k\gg 1$ the spectral density of the Laplacian Eq.(\ref{Eq:LapDef}) is given \cite{Kahng07} as
\beqarr
	\rho(\lambda) &=& 	\left\lbrace 
					\begin{array}{l l}
						(\alpha-1)(-\lambda_c)^{\alpha-1}~(-\lambda)^{-\alpha}	
						& \quad \textnormal{for }\lambda < \lambda_c < 0	~,	\\
						0 & \quad \textnormal{otherwise}	~,
					\end{array}
				\right.		\nonumber \\ \\
	\lambda_c &=& -k(\alpha-2)(\alpha-1)^{-1}	~.		\nonumber 
\eeqarr
Using this spectral density in equation Eq.(\ref{Eq:SpecAv}) we obtain
\beq
	\mathbb{E}\left[\Omega^{(2)}\right] ~=~ \gamma \int d\lambda ~\rho(\lambda)~\frac{1}{\lambda} ~=~ \gamma k^{-1} \frac{(\alpha-1)^2}{\alpha(\alpha-2)}	~.
\eeq
\subsection*{Erd\H{o}s-R\'enyi model and random tree network limit}
\noindent
The Erd\H{o}s-R\'enyi random model \cite{REgraph59} is recovered as a special case of the static scale-free random network model in the limit $\alpha\to\infty$ \cite{Kahng01}. Then for $k\gg 1$ we find $\mathbb{E}\left[\Omega^{(2)}\right]=\gamma k^{-1}$. 
One can study the limit of sparse uncorrelated random graphs by removing edges randomly without breaking the network in two components. The mean degree in a single component, undirected graph cannot be smaller than $2(N-1)/N$ for a tree network. Every edge that is removed then breaks the connectivity, creates a new component and thus a new zero eigenvalue of the Laplacian. We expect therefore a divergence of the integral in Eq. (\ref{Eq:SpecAv}) for $k\to 2$.
We have tested our theory numerically for symmetrically connected random graphs with Poissonian uncorrelated degree distributions and $N=400$ phase oscillators with nonisochronicity $\gamma=1.0$. The frequencies were chosen randomly from a uniform distribution with $\textnormal{var}(\omega)=\sigma^2=10^{-2}$. The synchronization frequency $\Omega$ was determined on one hand by solving the algebraic equations Eq.(\ref{Eq:KPESynch01}) with a Newton method, instead of integrating the KPEs Eq.(\ref{Eq:KPE01}), and on the other hand by using our perturbation approach and the complete eigenvalue spectrum of the network Laplacians. The results can be seen in Fig.~(\ref{Fig:fig01}). One can, indeed, see that the second order perturbation term diverges as $(k-2)^{-1}$ which is consistent with a powerlaw scaling $\mathbb{E}\left[\Omega^{(2)}\right]=\gamma k^{-1}$ for larger mean degrees $k$.
\subsection*{Application to network structure analysis}
\noindent
In order to demonstrate possible applications of this perturbation theory to structural analysis of an unknown coupling network let us now briefly study what information can be gained from a measurement of linear and nonlinear responses to frequency changes of the oscillators. In \cite{Timme07} the author presents a method to reconstruct a coupling network from measuring the linear response of the phase differences to linearly independent changes of the natural frequencies. This corresponds to using Eq.~(\ref{Eq:PertSol01}).
The coupling network can be identified from the Green's function $\mathbf{G}$ of the network Laplacian and Eq. (\ref{Eq:PertSol01}) reads
\beq
	\boldsymbol{\vartheta}^{(1)} = \mathbf{G}\boldsymbol{\eta}	~.
\eeq
If the phase differences are not accessible to direct measurement one can in principle also obtain the Green's function from the second order shift in the synchronization frequency. For a symmetric coupling Eq. (\ref{Eq:PertSol02}) gives
\beq
	\Omega(\boldsymbol{\omega}) = \bar\omega + \gamma~ \frac{1}{N} \boldsymbol{\omega}^\dagger \mathbf{G} \boldsymbol{\omega}	~.
\eeq
Let $\lbrace \boldsymbol{\omega}^{(\alpha)} \rbrace$ be a basis set of linear independent frequency detunings. Then the Green's function with respect to this basis can be determined from $N(N+1)/2$ measurements of synchronization frequencies as
\beq
	\Omega\left(\boldsymbol{\omega}^{(\alpha)}+\boldsymbol{\omega}^{(\beta)}\right) - \left(\Omega\left(\boldsymbol{\omega}^{(\alpha)}\right) + \Omega\left(\boldsymbol{\omega}^{(\beta)}\right)\right) 
	= 2\gamma \frac{1}{N} ~\boldsymbol{\omega}^{(\alpha)\dagger} \mathbf{G} \boldsymbol{\omega}^{(\beta)}	~.
\eeq
However, due to the number of measurements and the large time scales of a diffusion process the application is limited to very small networks, fast relaxation to 
the phase locked solution
and high precision measurements. The analysis can be extended to nonidentical oscillators and asymmetric coupling.
\section*{V. Discussion}
\noindent
We have presented expressions for the first and second order perturbation terms of the synchronization frequency in complex networks of coupled Kuramoto phase oscillators with quenched frequency disorder. The two approaches in Sections II and III give equivalent results, but the second approach, based on a nonlinear approximation of the phase coupling function around zero, extends a well known treatment of the Kuramoto phase equations from continuous media to complex networks \cite{Kuramoto84,Sakaguchi88,BlaToe08}. The results were given in terms of the eigenvalues and eigenvectors of the Laplacian matrix of the coupling network. In a single component with mean degree $k$ of a 
%symmetric 
undirected
Erd\H{o}s-R\'enyi random coupling network \cite{REgraph59} and for oscillators with independent, identically distributed random frequencies $\omega_n$ of variance $\sigma^2$ and nonisochronicity $\gamma$ the expected synchronization frequency was found to be
\beq
	\mathbb{E}\left[\Omega\right] ~=~ \mathbb{E}\left[\omega\right] ~+~ \gamma~\textnormal{var}(\omega)~(k-2)^{-1} ~+~ O(\gamma^2\sigma^3)	~.
\eeq
While the expected synchronization frequency depends to the first order only on the natural frequencies in the system, the second order correction combines the nonlinearity $\gamma$ of the phase coupling function around zero, the variance of the frequencies and the mean degree of the coupling network in a simple way. 
\\ \\
The explicit connection between synchronization frequency, natural frequencies and network structure in the Eq.~(\ref{Eq:PertSol02}) makes it in principle possible to infer information of either property from a measurement or the knowledge of the other properties. Network reconstruction by observing the linear response to a frequency detuning has already been proposed and successfully applied \cite{Timme07}. An analogous approach using frequency measurements instead of phase differences may be constructed based on the results of this paper.
\\ \\
This work was supported by the German DFG through the project SfB555 and the Japanese JSPS.
\section*{Appendix A}
\noindent
Given the Kuramoto phase equations in synchronization
\beq	\label{Eq:KPESynchApp}
	\Omega = \sigma\eta_n + \sum_{m=1}^N A_{nm}~g(\vartheta_m-\vartheta_n)	~,
\eeq
and a phase locked solution of the KPEs for identical oscillators
\beq	\label{Eq:KPEHomSynch}
	\Omega^{(0)} = \sum_{m=1}^N A_{nm}~g(\vartheta^{(0)}_m-\vartheta^{(0)}_n)	~,
\eeq
we want to derive expressions for the coefficients in the expansion of the synchronization frequency $\Omega$ in powers of the frequency heterogeneity $\sigma$. 
Let us start by implicitly defining notations for the involved perturbation terms and phase differences
\beqarr
	\Omega &=& \Omega^{(0)} + \sum_{l=1}^\infty \sigma^l~\Omega^{(l)} 	~,	\label{Eq:OhmExpansion} \\ \nonumber \\
	\vartheta_n &=& \vartheta_n^{(0)} + \varphi_n ~=~ \vartheta_n^{(0)} + \sum_{l=1}^\infty \sigma^l ~\vartheta_n^{(l)}	~,	\label{Eq:ThExpansion} \\ \nonumber \\
	\vartheta_{mn} &=& \vartheta_m - \vartheta_n ~=~ \vartheta_{mn}^{(0)} + \varphi_{mn} ~=~ \vartheta_{mn}^{(0)} + \sum_{l=1}\sigma^l~\vartheta_{mn}^{(l)}		~,	\label{Eq:ThDiffDef} 
	\\ \nonumber \\
	\varphi_{mn}^j &=& (\varphi_m - \varphi_n)^j = \sum_{l=j}^\infty \sigma^l ~\varphi_{mn}^{(j,l)} 	~,	\label{Eq:PhiPowExpansion}\\ \nonumber \\
	g_{mn} &=& g\left(\vartheta_{mn}^{(0)}\right) ~,\qquad g_{mn}^{(j)} = \partial^j g_{mn}	\label{Eq:GmnDef}	~.
\eeqarr
Note that here we do not assume $\boldsymbol{\vartheta}^{(0)}=\textnormal{const}$, $g(0)=0$ or $g_{mn}^{(j)}=g^{(j)}(0)$. It has been pointed out, that even for identical oscillators the homogeneous solution may not be the only synchronized solution of the Kuramoto phase equations \cite{Strogatz06}. In certain coupling topologies and for large nonisochronicity the completely synchronized solution can coexist with a dominating chaotic attractor of drifting phases \cite{ErmChimera08}. If the network is homogeneous and sufficiently well connected, however, the stable solution of complete synchronization is typical. Therefore we assume $\Omega^{(0)}=g(0)=0$ and $g^{(j)}_{mn}=g^{(j)}(0)$ in the main text of this paper.
\\ \\
The coefficients $\varphi_{mn}^{(j,l)}$ yield the recursion relation
\beq	\label{Eq:RecursionRel}
\varphi_{mn}^{(j,l)} = \left\lbrace
			\begin{array}{l l}
				\vartheta_{mn}^{(l)} & ~~\textnormal{for} \quad j=1 	~,	\\
				\sum_{k=1}^{l-1} \vartheta_{mn}^{(k)}\varphi_{mn}^{(j-1,l-k)} & ~~\textnormal{for}\quad j\le l 	~,\\
				0  & ~~\textnormal{otherwise}	~.
			\end{array}
		      \right.
\eeq
Inserting Eq.(\ref{Eq:ThExpansion}) into Eq.(\ref{Eq:KPESynchApp}) we find
\beqarr
	\Omega 	&=& \sigma\eta_n + \sum_{m=1}^N A_{nm}~g_{mn}	~,	\\ \nonumber \\
		&=& \Omega^{(0)} + \sigma\eta_n + \sum_{m=1}^N A_{nm}~\sum_{j=1}^\infty \frac{1}{j!}g_{mn}^{(j)} \varphi_{mn}^j 	~,	\nonumber \\ \nonumber \\
		&=& \Omega^{(0)} + \sigma\eta_n + \sum_{m=1}^N A_{nm}~\sum_{j=1}^\infty \frac{1}{j!}g_{mn}^{(j)} \sum_{l=j}^\infty \sigma^l \varphi_{mn}^{(j,l)} 	~,	\nonumber \\ \nonumber \\
		&=& \Omega^{(0)} + \sigma\eta_n + \sum_{l=1}^\infty \sigma^l \sum_{j=1}^l  \sum_{m=1}^N A_{nm}~\frac{1}{j!}g_{mn}^{(j)} 
			\sum_{k=1}^{l-1} \varphi_{mn}^{(1,k)}\varphi_{mn}^{(j-1,l-k)} ~. \nonumber
\eeqarr
In the second line we have inserted the unperturbed solution Eq.(\ref{Eq:KPEHomSynch}) and in the third line we used the expansion Eq.(\ref{Eq:PhiPowExpansion}) of the powers of $\varphi_{mn}$. Since the leading order of $\varphi_{mn}$ in $\sigma$ is one, the $j$th power has a leading term of order $j$. In the last line the recursion relation (\ref{Eq:RecursionRel}) was used. If we now collect the nonlinear terms ($j>1$) in a vector $\mathbf{b}^{(l)}$ we can write down this result in a more compact form
\beq
	\Omega-\Omega^{(0)} = \sum_{l=1} \sigma^l \Omega^{(l)} = \sum_{l=1} \sigma^{l}  \left( \sum_{m=1}^N A_{nm}g'_{mn} \vartheta_{mn}^{(l)} + b_n^{(l)} \right)	~,
\eeq
or in vector form 
\beq	\label{Eq:OhmVecExp}	
	\Omega^{(l)} \mathbf{1} = \left( \mathbf{L} \boldsymbol{\vartheta}^{(l)} + \mathbf{b}^{(l)} \right)	~,
\eeq
where $\mathbf{1}$ is a constant vector of unit entries, $\boldsymbol{\vartheta}^{(l)}$ is the vector of perturbative corrections $\vartheta_n^{(l)}$ to the phases
and $\mathbf{b}^{(l)}$ is a vector which only depends on perturbation terms of order lower than $l$.
The matrix $\mathbf{L}$ is the Jacobian
\beq	\label{Eq:LapDefApp}
	L_{nm} = A_{nm}~g'_{mn} - \delta_{nm}\sum_{l=1}^N A_{nl}~g'_{ln}	~,
\eeq
or the Laplacian if $g'_{mn}$ is a constant. The vectors $\mathbf{b}^{(l)}$ are given by
\beqarr
	b_n^{(1)} &=& \eta_n	~,	\\ \nonumber \\
	b_n^{(2)} &=& \sum_{m=1}^N A_{nm}~\frac{1}{2}g_{mn}''{\vartheta_{mn}^{(1)}}^2 	~,	\\ \nonumber \\
	b_n^{(3)} &=& \sum_{m=1}^N A_{nm} \left( 	g_{mn}''\vartheta_{mn}^{(1)}\vartheta_{mn}^{(2)} + \frac{1}{6}g_{mn}'''{\vartheta_{mn}^{(1)}}^3 \right)		~,
\eeqarr
and in general for $l>1$
\beqarr
	b_n^{(l>1)} = \sum_{j=2}^l \sum_{m=1}^N A_{nm} \frac{1}{j!}g_{mn}^{(j)} \sum_{k=1}^{l-1}\vartheta_{mn}^{(k)}\varphi_{mn}^{(j-1,l-k)}	~.
\eeqarr
Equation Eq.(\ref{Eq:OhmVecExp}) can be solved iteratively for each perturbation order.
Let us consider a complete, orthonormal set of left and right eigenvectors $\mathbf{P}_k$ and $\mathbf{p}_k$ of the Jacobian with
\beqarr
	\mathbf{L}\mathbf{p}_k = \lambda_k \mathbf{p}_k ~,&&\quad \mathbf{L}^\dagger\mathbf{P}_k = \lambda_k^*\mathbf{P}_k 	~,\nonumber \\ \\
	\mathbf{P}_k^\dagger{~}\mathbf{p}_{k'} = \delta_{kk'} ~,&&\quad \sum_{k=0}^{N-1} \mathbf{p}_k\mathbf{P}_k^\dagger = \mathbb{I}	~,	\nonumber
\eeqarr
and in particular
\beq
	\mathbf{p}_0 = \mathbf{1}	~,\qquad		\mathbf{1}^\dagger{~}\mathbf{P}_0 = 1		~.
\eeq
We can now define the projectors 
\beq
	\mathbb{P}_0 = \mathbf{p}_0\mathbf{P}_0^\dagger	~,	\qquad		\mathbb{Q}_0 = \mathbb{I}-\mathbb{P}_0		~.
\eeq
The operation $\mathbb{P}_0\mathbf{x}$ projects to a constant vector where all entries are equal to the weighted average $\langle x\rangle_{\mathbf{P}_0}$ and $\mathbb{Q}_0$ removes this average from the components of a vector. Applying these projectors to Equation Eq.(\ref{Eq:OhmVecExp}) we obtain
\beqarr
	\Omega^{(l)} &=& \mathbf{P}_0^\dagger{~} \mathbf{b}^{(l)} 	~, \\ \nonumber \\
	0 &=& \mathbf{L}\boldsymbol{\vartheta}^{(l)} + \mathbb{Q}_0 \mathbf{b}^{(l)} 	~.		\label{Eq:PertVecProblem01}
\eeqarr
The last equation is solved for $\boldsymbol{\vartheta}^{(l)}$ up to an arbitrary global phase shift by
\beq
	\boldsymbol{\vartheta}^{(l)} = -\sum_{k\ne 0} \frac{ \left(\mathbf{P}_k^\dagger{~}\mathbf{b}^{(l)}\right) }{\lambda_k} \mathbf{p}_k	~.
\eeq
%
%
%
%
%\clearpage
%
%
%
%
%
%

\end{document}